\documentclass[a4paper]{jpconf}
\usepackage{graphicx}
\usepackage{hyperref}
\usepackage{amssymb}
\usepackage{amsmath}

\def\bea {\begin{eqnarray}}
\def\eea {\end{eqnarray}}
\def\ra {\rightarrow}
\def\be {\begin{equation}}
\def\ee {\end{equation}}
\def \beq{\begin{equation}}
\def \eeq{\end{equation}}
\def \beqa{\begin{eqnarray}}
\def \eeqa{\end{eqnarray}}
\def \la{\langle}
\def \ra{\rangle}
\def \l{\left(}
\def \r{\right)}
\def \l{\left(}
\def \r{\right)}

\newcommand{\pT}{$p_{\rm{T}}$}
\newcommand{\kT}{$k_{\rm{T}}$}

\newcommand{\muB}{$\mu_{\rm B}$}

\newcommand{\sNN}{$\sqrt{s_{\rm NN}}$}
\newcommand{\Ss}{$S\sigma$}
\newcommand{\KV}{$\kappa \sigma^2$}

\newcommand{\cv}{$c_{\rm v}$}

\newcommand{\meanpt}{$\langle p_{\rm{T}} \rangle$}

\newcommand{\MV}{$\sigma^{2}/M$}

\begin{document}
\title{Probing the QCD phase structure using event-by-event fluctuations}

\author{Tapan K. Nayak}

\address{CERN, CH-1211, Geneva 23, Switzerland, and \\
National Institute of Science Education and Research, HBNI, Bhubaneswar, India}

\ead{Tapan.Nayak@cern.ch}

\begin{abstract}
Heavy-ion collisions at relativistic energies probe matter at extreme
conditions of temperatures and energy densities. 
The study of event-by-event fluctuations of experimental observables is
crucial to probe the QCD phase transition, locate the critical point,
and learn about the associated critical phenomena. At the critical
point, all thermodynamic quantities  
behave anomalously. Fluctuation measurements provide access to
thermodynamic response functions. We discuss the methods for obtaining
the isothermal compressibility using particle multiplicity
fluctuation, and specific heat using fluctuations in mean transverse momentum, temperature, and 
energy. Lattice QCD calculations have predicted non-monotonic behavior
in the higher-order cumulants of conserved quantities at the critical
point. Fluctuations in the multiplicity of
charged to neutral particles have been measured to understand 
the formation of domains of disoriented
chiral condensates. 
We review the recent fluctuation results as a function of
collision centrality and energy from experiments
at SPS, RHIC, and LHC. 
In addition, we propose to map the temperature fluctuations in 
$\eta$-$\phi$ plane to probe local fluctuations of temperature and
energy density.
\end{abstract}

\section{Introduction}

\begin{figure}[tbp]
\begin{center}
\includegraphics[width=16pc,height=15pc]{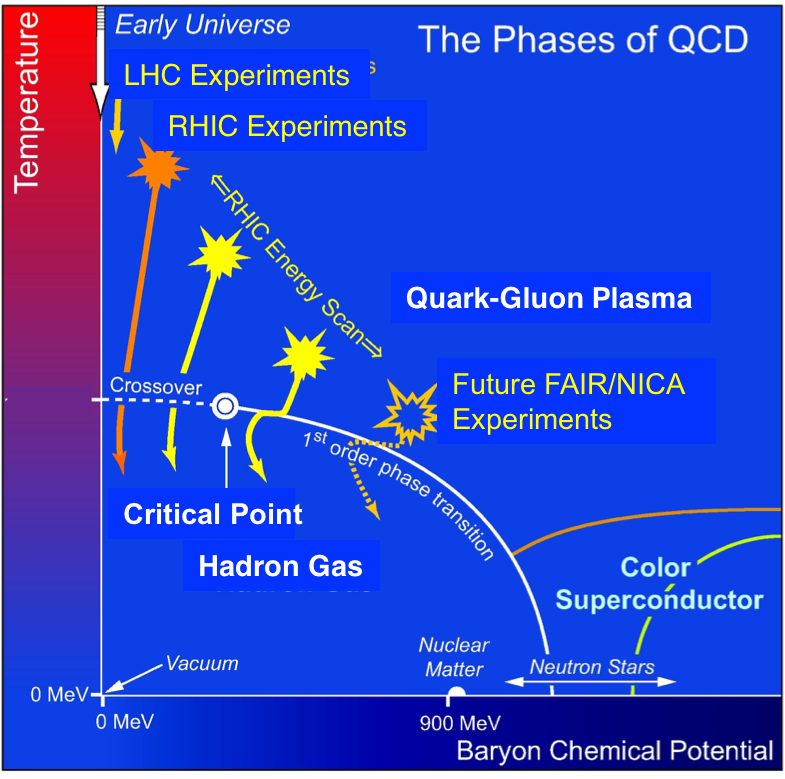}
\caption{\label{phases} 
A schematic QCD phase diagram in the temperature ($T$) and baryonic 
chemical potential ($\mu_B$) plane. The regions probed by 
different accelerator facilities are indicated.}
\end{center}
\end{figure}

According to the theory of Quantum Chromo Dynamics (QCD), under extreme
conditions of  temperatures and energy densities, normal hadronic matter goes
through a phase transition to a system of deconfined quarks and
gluons, the quark-gluon plasma (QGP). The QCD phase structure between
these two distinct states of matter span a wide range of baryon chemical potential
($\mu_{\rm B}$) and temperature ($T$) as shown in Fig.~\ref{phases}.
Lattice QCD calculations indicate that at vanishing \muB, the
transition from the QGP to a hadron gas is a smooth 
crossover~\cite{lattice1,lattice2,lattice3,lattice4}, while at large \muB, the phase transition is
of first order~\cite{hg1,hg2}.
The point in $T$ and \muB~where the first order transition ends and
instigates a crossover transition is denoted as the QCD critical
point. Theoretical and experimental
studies explore the rich landscape of the QCD phase diagram to
understand the nature of the phase transition, locate the critical point, and to learn about the
properties of the matter formed. 

The experimental program to study the QCD phase structure
started more than three decades ago at 
Bevelac, Berkeley and since then has covered four generations of
experiments at the Brookhaven National Laboratory (BNL) and
CERN. The collisions at the Large Hadron Collider (LHC) and at the top energy at the
Relativistic Heavy Ion Collider (RHIC) probe the conditions at low
\muB.  The beam energy scan (BES) program
at RHIC~\cite{bes1,bes2} is specially designed to probe 
the location of the critical point
by varying the collision energy at close intervals in $T$-$\mu_{\rm B}$. In future, this
program will be complemented by upcoming facilities at Dubna,
Russia, and GSI, Germany to explore large \muB~regions in the phase
diagram. The regions probed by different accelerator
facilities are indicated in Fig.~\ref{phases}. 

Fluctuations play a crucial role in the study of phase
transition and any associated critical phenomena. 
Event-by-event fluctuations in a number of observables
have been predicted as signatures of the QCD phase transition and the
critical point~\cite{stephanov1,heiselberg,karsch}. 
Several thermodynamic quantities show varying fluctuation
patterns when the system goes through the phase boundary. 
The defining
characteristics of the QCD phase transition are the abrupt changes in
the physical properties of the system which can be inferred through
the analysis of fluctuations of different observables. At the critical
point, the fluctuations are expected to be very large.
The main fluctuation signatures emanate in the form of event-by-event
measurement in the number of particles, momenta of particles, 
as well as the spatial and energy driven patterns of multiplicity
distributions. 
In this article, we discuss some of the fluctuation
techniques, experimental results, and future prospects.
\begin{itemize}
\item{Thermodynamic response functions: Response functions such as  
isothermal compressibility (\kT), specific heat (\cv), and speed of 
sound, are related by the equation of 
    state (EOS), which governs the evolution of the system. The nature 
    of phase transitions in a system can be understood by the measurement of 
    thermodynamic response functions. 
    These quantities can be accessed experimentally by the fluctuation of 
    measured quantities. The heat capacity is related to the 
    fluctuations in temperature~\cite{landau,sto}, whereas in the 
    grand canonical ensemble (GCE) framework  \kT~is related 
    to the fluctuation in particle multiplicity~\cite{mrow}. 
    Skewness of mean 
    transverse momentum (\meanpt) fluctuations has recently been proposed 
    as a probe of hydrodynamic behavior in nuclear collisions~\cite{jaki}.
    By measuring the event-by-event fluctuations in particle 
    multiplicity ($N$), \meanpt, and mean transverse energy, we can 
    get access to the response functions. 
}
\item{Fluctuations of conserved quantities:
Lattice QCD calculations reveal that
the higher order cumulants of conserved quantities, such as net-charge ($Q$), net-proton ($B$), and
net-strangeness ($S$), within a limited acceptance, are
proportional to the powers of the correlation length and are expected
to diverge at the critical point~\cite{stephanov2,stephanov3}. 
Experimentally, it is possible to measure $Q$, $B$, and $S$ on an
event-by-event basis and obtain the cumulants of these distributions.
In addition, off-diagonal cumulants explore the flavor
carrying susceptibilities of the system~\cite{koch,majumdar,gavai,arghya}.
Assuming that the signal at freeze-out survives dissipation during the 
evolution of the fireball from the hadronization stage, the higher 
cumulants can be used as one of the preferred tools for locating the 
critical point. }

\item{Disoriented chiral condensates (DCC): 
    Disoriented chiral condensates (DCC), localized in phase space, 
    have been predicted to be formed in high energy heavy-ion
    collisions when the chiral symmetry is restored at high temperatures~\cite{bjorken,rajagopal}. 
    Anomalous event-by-event fluctuations of the neutral to charged pions as
    well as neutral to charged kaons have been predicted as signatures
    of the formation of DCC. A 
    fresh look at RHIC and LHC energies is needed to infer about the formation of the DCC domains. 
}
\item{Fluctuation map: The physics of heavy-ion 
collisions at ultra-relativistic energies has often been compared to the Big Bang 
phenomenon of the early Universe. 
Observation of the cosmic microwave background radiation (CMBR) by various satellites confirms the Big 
Bang evolution, and inflation; that provides important information regarding the early Universe and its evolution 
with excellent accuracy. The matter produced at extreme conditions of energy density and 
temperature in heavy-ion collisions is a Big Bang 
replica on a tiny scale ~\cite{heinz,hannu,urs}. 
We propose to map the temperature fluctuations in 
$\eta$-$\phi$ plane to probe local fluctuations of temperature and
energy density.
}
\end{itemize}

\section{Multiplicity fluctuation and estimation of isothermal compressibility}

Isothermal compressibility (\kT) is the measure of
the relative change in volume with respect to change in
pressure~\cite{mrow},
\beqa
\left.k_T\right|_{T,\la N\ra} &=&
-\frac{1}{V}\left.\l\frac{\partial V}{\partial P}\r\right|_{T,\la N\ra}
\label{eq.kT}
\eeqa
where $V, T, P$ represent volume, temperature, and pressure of the
system, respectively, and $\la N\ra$ stands for the mean yield of the particles. 
In the Grand Canonical Ensemble (GCE) framework, the variance of the number of particles
is directly related to \kT:
\beqa 
\omega_{\rm ch} = \frac{k_{\rm B}T \la N\ra}{V}k_{\rm T},
\label{imp}
\eeqa
where $\omega_{\rm ch} $ is the scaled variance of multiplicity distribution.
This formalism may be applied to experimental measurements of multiplicity at 
mid-rapidity, as energy and conserved quantum numbers are exchanged 
with the rest of the system.  

\begin{figure}[h]
\begin{center}
\begin{minipage}{16pc}
\includegraphics[width=16pc,height=12pc]{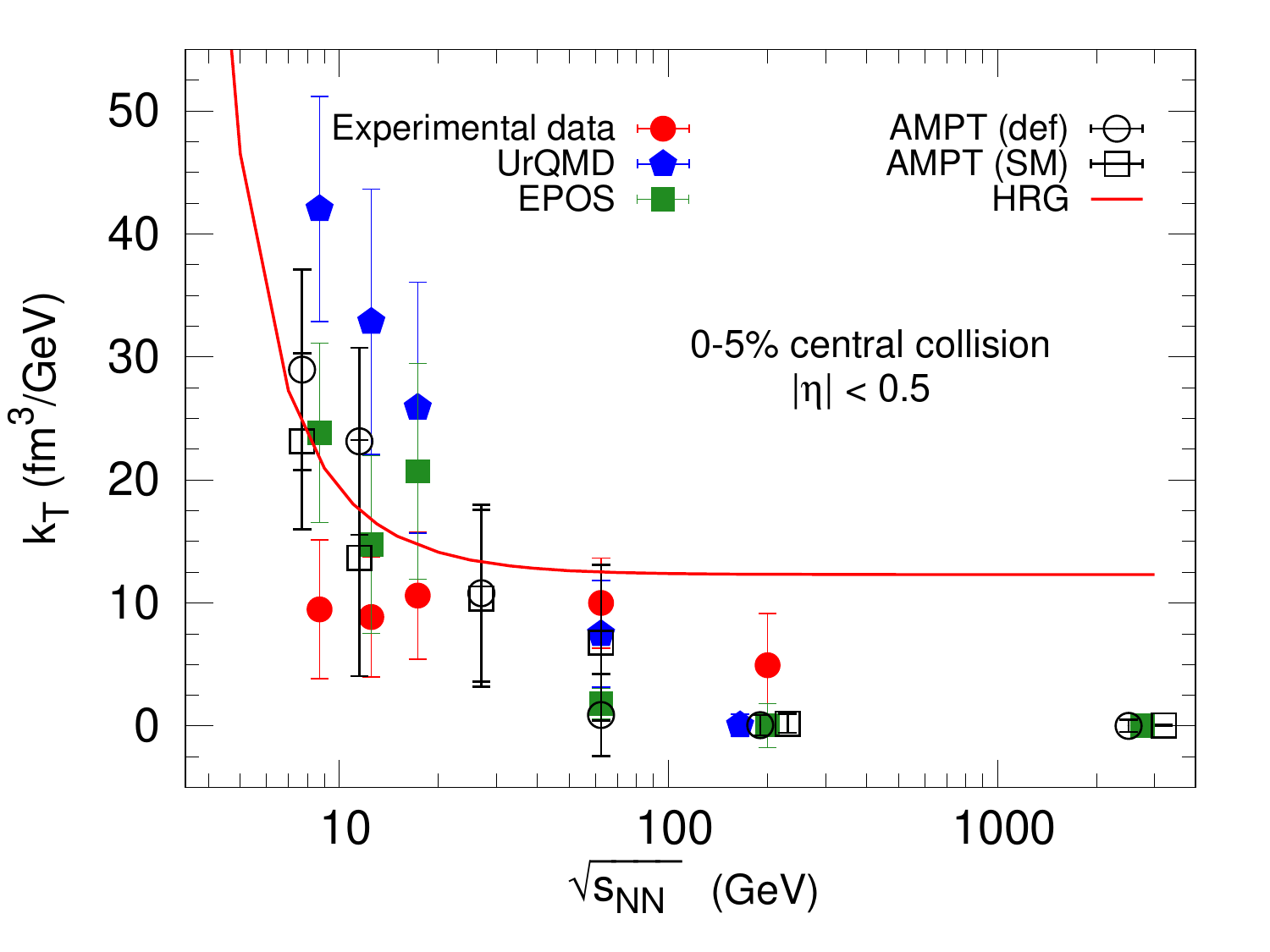}
\vspace{-1pc}\caption{\label{compressibility}
Isothermal compressibility, $k_{\rm T}$ as a function of collision energy 
for central collisions~\cite{mait}.}
\end{minipage}\hspace{2pc}%
\begin{minipage}{16pc}
\vspace{0pc}%
\includegraphics[width=15pc,height=11pc]{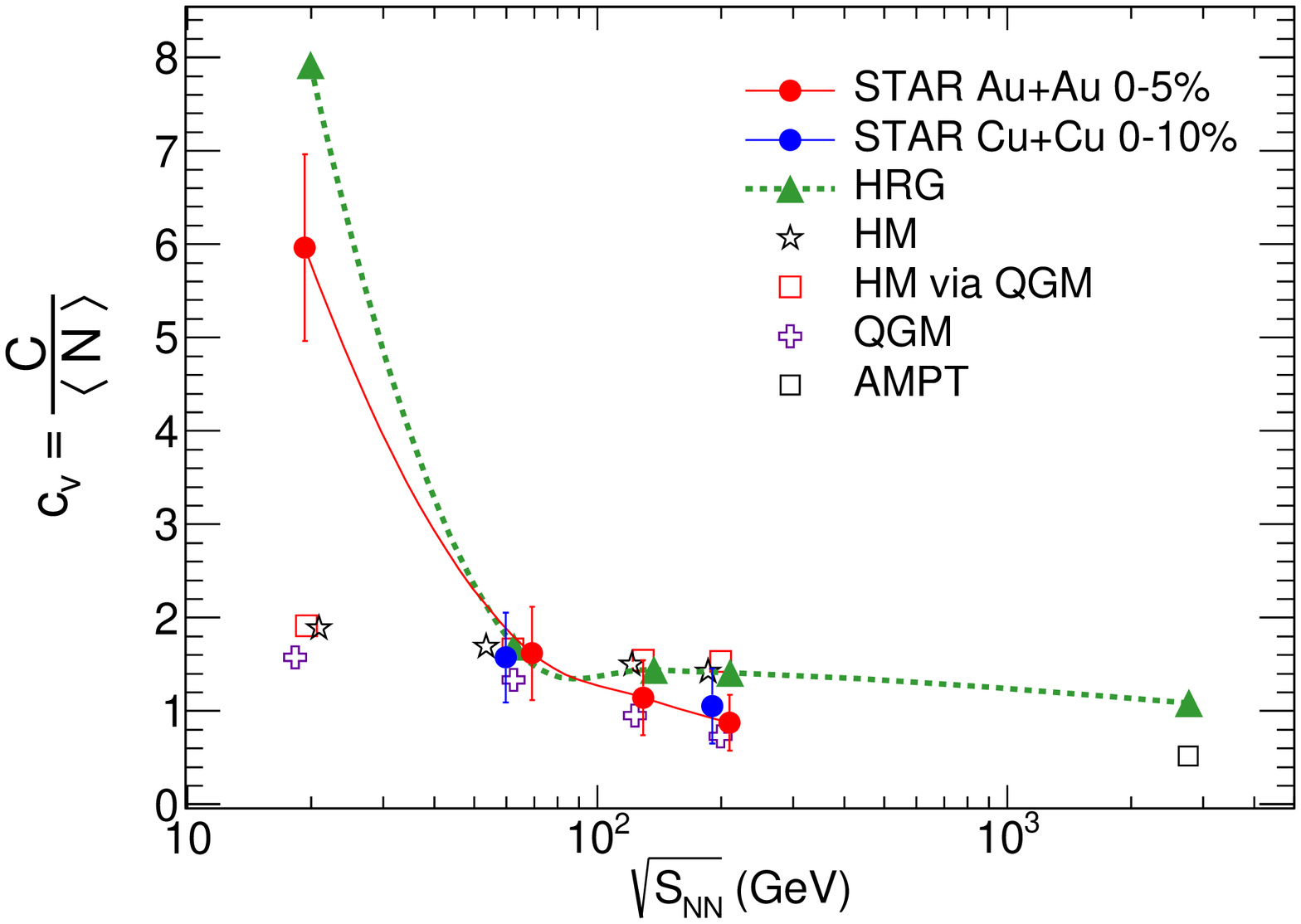}
\vspace{1pc}\caption{\label{specific_heat} 
Specific heat, $c_{\rm v}$ as a function of collision energy 
for central collisions~\cite{sumit}.}
\end{minipage} 
\end{center}
\end{figure}

At the chemical freeze-out surface, the inelastic collisions cease,
and thus the hadron multiplicities get frozen.  While the ensemble average
thermodynamic properties like the temperature and volume can be
extracted from the mean hadron yields, \kT~can be accessed through
the measurements of the event-by-event multiplicity fluctuations. This
has been explored in Ref.~\cite{mait} and the estimated values of
\kT~are shown in Fig.~\ref{compressibility} for available experimental
data and estimations from the HRG calculations. The HRG calculations reveal a sharp increase in the value of
\kT~around \sNN$\sim$20~GeV. It will be interesting to explore the
behavior of \kT~at BES-II energies and future facilities.

\section{Temperature and $\langle p_{\rm T}\rangle$ fluctuations and estimation of specific heat}

Heat capacity ($C$) is a response function which expresses how much
the temperature of a system changes when the heat is transferred to 
it. The
specific heat (\cv) is heat capacity divided by the number of 
charged particles. 
Equivalently, for a system in thermal 
equilibrium to a bath at temperature, $T$, we can write~\cite{landau,sto}:
\begin{eqnarray}
C = \biggl(  \frac{\partial E}{\partial T} \biggr)_{V} 
= \frac{(\langle E^2 \rangle -  \langle E \rangle ^2 )}  
    { \langle T \rangle ^2}. 
\label{eqn1}
\end{eqnarray}
where $V$ and $E$ are volume and energy of the system,
respectively. By expressing the transverse momentum spectra of 
emitted particles in terms of event-by-event temperature fluctuation ($\Delta T = T - \langle T \rangle$),
we obtain:
\begin{eqnarray}
P(T) \sim \exp [-\frac{C}{2}  \frac{(\Delta T)^2}{\langle T 
  \rangle^2}], ~~~~~~~~ \frac{1}{C} = 
\frac{(\langle T^2 \rangle -  \langle T \rangle ^2 )}  
   { \langle T \rangle ^2}. 
\label{eqn3}
\end{eqnarray}
Heat capacity thus can be estimated from the fluctuations in energy or temperature. 
For a system in equilibrium, the mean values of $T$ and $E$ are 
related by an equation of state. However, the fluctuations in energy 
and temperature have very different behavior. 
Energy being an extensive quantity, its 
fluctuation has a volume dependent component. So energy is 
not suited for obtaining the heat 
capacity. On the other hand, temperature fluctuations provide a good 
major for the estimation of specific heat. 
The temperature of 
the system can be obtained from the transverse momentum (\pT) spectra of the 
emitted particles. For a range of \pT~within $a$ to $b$, we obtain~\cite{sumit}:
\begin{eqnarray}
\langle p_{\rm T} \rangle 
&=& 
\frac 
{\int_a^b p_{\rm T}^2  F(p_{\rm T})  dp_{\rm T} }
{\int_a^b p_{\rm T} F(p_{\rm T})  dp_{\rm T} } 
=  
2T_{\rm eff} +  
  \frac 
{a^2    e^{-a/T_{\rm eff}} -     b^2 e^{-b/T_{\rm eff}}   }
{(a+T_{\rm eff}) e^{-a/T_{\rm eff}} - (b+T_{\rm eff}) e^{-b/T_{\rm eff}}   }. 
\label{eqn7}
\end{eqnarray}
This equation links the value of event-by-event \meanpt~within a 
specified range of \pT~to the apparent or effective temperature 
($T_{\rm eff}$). The dynamical component of the \meanpt~and $T$ distributions 
have been obtained for RHIC energies after subtracting the 
corresponding mixed event distributions. 
The heat capacity is calculated using the 
dynamical part of the fluctuation and the kinetic temperature. 
In Ref.~\cite{sumit}, $\langle p_{\rm T} \rangle $ results from the
STAR collaboration have been used to calculate heat capacity.
The results of the specific heat as a function of collision energy are presented in 
Fig.~\ref{specific_heat}. It shows a sharp rise in \cv~below 
\sNN~=~62.4~GeV. 
In order to probe the QCD critical point, a finer scan of beam energy at RHIC is essential.

\section{Dynamical net-charge fluctuations}

\begin{figure}[tbp]
\begin{center}
\begin{minipage}{17pc}
\includegraphics[width=17pc]{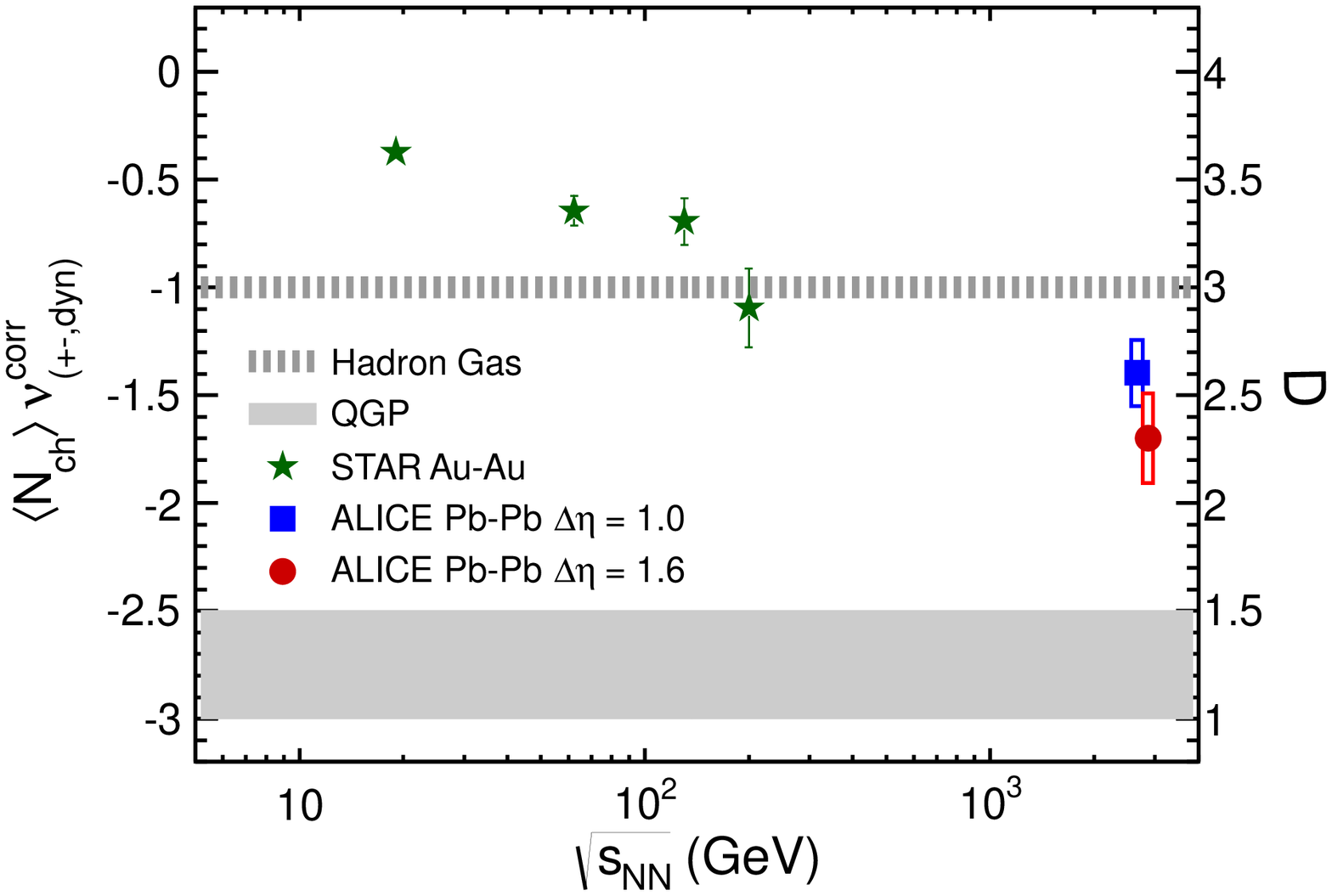}
\caption{\label{nudyn_1}Collision energy dependence of the net-charge 
fluctuations in central collisions~\cite{ALICE_nudyn}.}
\end{minipage} 
\hspace{2pc}%
\begin{minipage}{17pc}
\includegraphics[width=17pc]{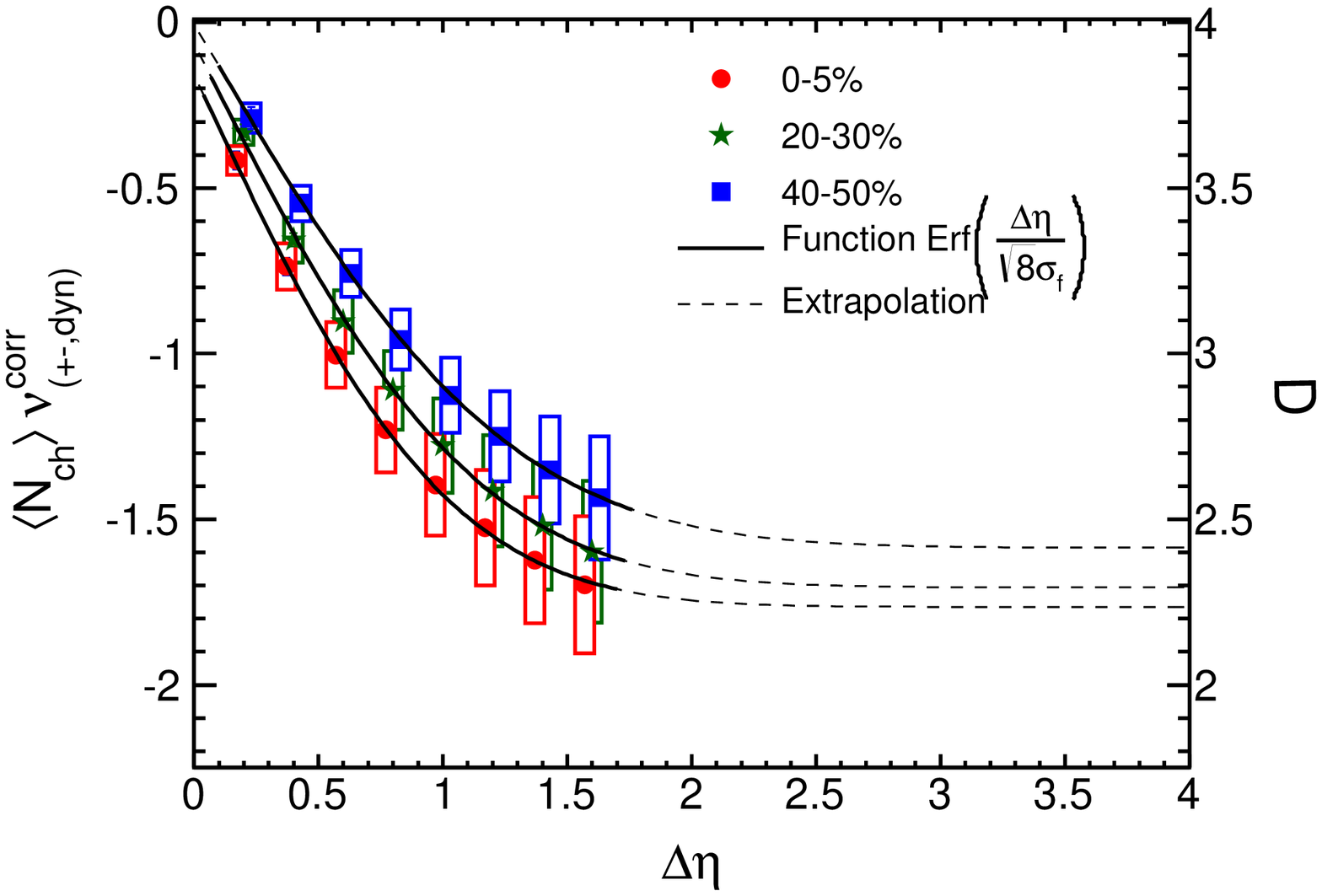}
\caption{\label{nudyn_2} Growth of net charge fluctuations as a 
  function the pseudorapidity window ($\Delta\eta$)~\cite{ALICE_nudyn}.}
\end{minipage}\hspace{2pc}%
\end{center}
\end{figure}

Net-charge fluctuations are strongly dependent on which phase the 
fluctuations originate. This is because, in the QGP phase, the charge carriers are quarks with fractional 
charges, whereas the particles in a hadron gas carry a unit 
charge. The fluctuations in the net charge depend on the squares of 
the charge states present in the system, and so, the net-charge 
fluctuations in the QGP phase are significantly smaller compared to 
that of a hadron gas. The initial QGP phase is 
strongly gluon dominated, and so the fluctuation per entropy may further be 
reduced as the hadronization of gluons increases the entropy. 
The charge fluctuations are best studied by calculating the quantity 
$\nu_{(+-,{\rm dyn})}$, defined as:
\begin{eqnarray}
\nu_{(+-,{\rm dyn})} = 
 \frac{\langle N_+(N_+-1) \rangle}{\langle N_+ \rangle ^2} +
 \frac{\langle N_-(N_--1) \rangle}{\langle N_- \rangle ^2} 
- 2\frac{\langle N_-N_+ \rangle}{\langle N_- \rangle \langle N_+ \rangle},
\end{eqnarray}
\noindent which is a measure of the relative correlation strength of 
particle pairs. 
A negative value of $\nu_{(+-,{\rm dyn})}$~signifies the dominant contribution from 
correlations between pairs of opposite charges. The $\nu_{(+-,{\rm 
    dyn})}$~has been found to be robust against random efficiency 
losses. It is related to the fluctuation measure ($D$) by:
\begin{eqnarray}
\langle N_{\rm ch} \rangle    \nu_{(+-,{\rm dyn})}  
\approx  D - 4, \\
{\rm where} ~~~~ D =  4 \frac{\langle \delta Q^2\rangle}{\langle N_{\rm ch} \rangle}. 
\label{Eq:D}
\end{eqnarray}
Here $Q=N_+-N_-$, and $N_{\rm   ch}=N_+-N_-$. 
Figure~\ref{nudyn_1} shows the values of $\langle N_{\rm 
  ch} \rangle    \nu_{(+-,{\rm dyn})}^{\rm corr}$ as a function of 
collision energy, measured by the STAR and ALICE 
Collaborations~\cite{ALICE_nudyn,STAR_nudyn}. Here $\nu_{(+-,{\rm dyn})}$ is corrected for 
global charge conservation and finite detector acceptance. A monotonic decrease in the value 
of $D$, measured has been observed. 

The measured fluctuations get diluted during the 
evolution of the system from hadronization 
to kinetic freeze-out because of the diffusion of charged hadrons in 
rapidity~\cite{pratt,shuryak1,aziz}. This has been studied by plotting 
dependence of net-charge fluctuations on the width of the rapidity 
window as shown in Fig.~\ref{nudyn_2} for Pb--Pb collisions at 
\sNN=2.76~TeV. 
We observe that for a given centrality bin, the 
$D$--measure shows a strong decreasing trend with the 
increase of $\Delta\eta$. In fact, 
the curvature of $D$ has a decreasing slope with a flattening 
tendency at large $\Delta\eta$~windows. 
The data points are fitted with a functional form, 
erf($\Delta\eta/\sqrt{8}\sigma_{\rm f}$), 
which represents the diffusion in rapidity space. The value of 
$\sigma_{\rm f}$, which represents the diffusion at freeze-out, turns 
out to be $0.41 \pm 0.05$ for central collisions. 
Taking the dissipation into account, 
the asymptotic value of fluctuations may represent the primordial 
fluctuations. It would be intriguing to make this study for a wider 
$\Delta\eta$ range in which most of the fluctuations are captured~\cite{bhanu}.

\section{Fluctuations of conserved quantities: diagonal cumulants}

Lattice QCD calculations have 
shown that higher order cumulants of the distributions of conserved 
charges are 
related to the corresponding higher-order thermodynamic susceptibilities and to the 
correlation length~($\xi$) of the system. These cumulants 
go through rapid changes near the critical point. 
Thus the measurement of cumulants of $Q$, $B$, and $S$ provide a direct correspondence to 
the lattice calculations and serve as an important probe for the 
critical point search. 

The event-by-event distributions of the conserved quantities within a 
limited acceptance are characterized by mean ($M$), standard deviation ($\sigma$), 
skewness ($S$), kurtosis 
($\kappa$), and other higher order cumulants. The products of the 
moments, such as \MV, \Ss, and \KV are constructed to cancel the volume 
term. The net-charge multiplicity distributions directly probe the 
charge quantum number, whereas the net-proton and net-kaon distributions 
provide good proxies for the net-baryon and net-strangeness conserved 
quantities. 

The STAR Collaboration has made extensive measurements of the cumulants 
of net-charge, net-kaon, and net-proton multiplicity distributions in 
Au--Au collisions at a wide range of energies starting from 
\sNN=7.7~GeV to 200~GeV. The collision centrality and energy 
dependence of the products of moments have been reported in a series 
of publications. Here we reproduce some of the salient features of the 
recent results. 

\begin{figure}[tbp]
\begin{center}
\begin{minipage}{16pc}
\includegraphics[width=16pc,height=18pc]{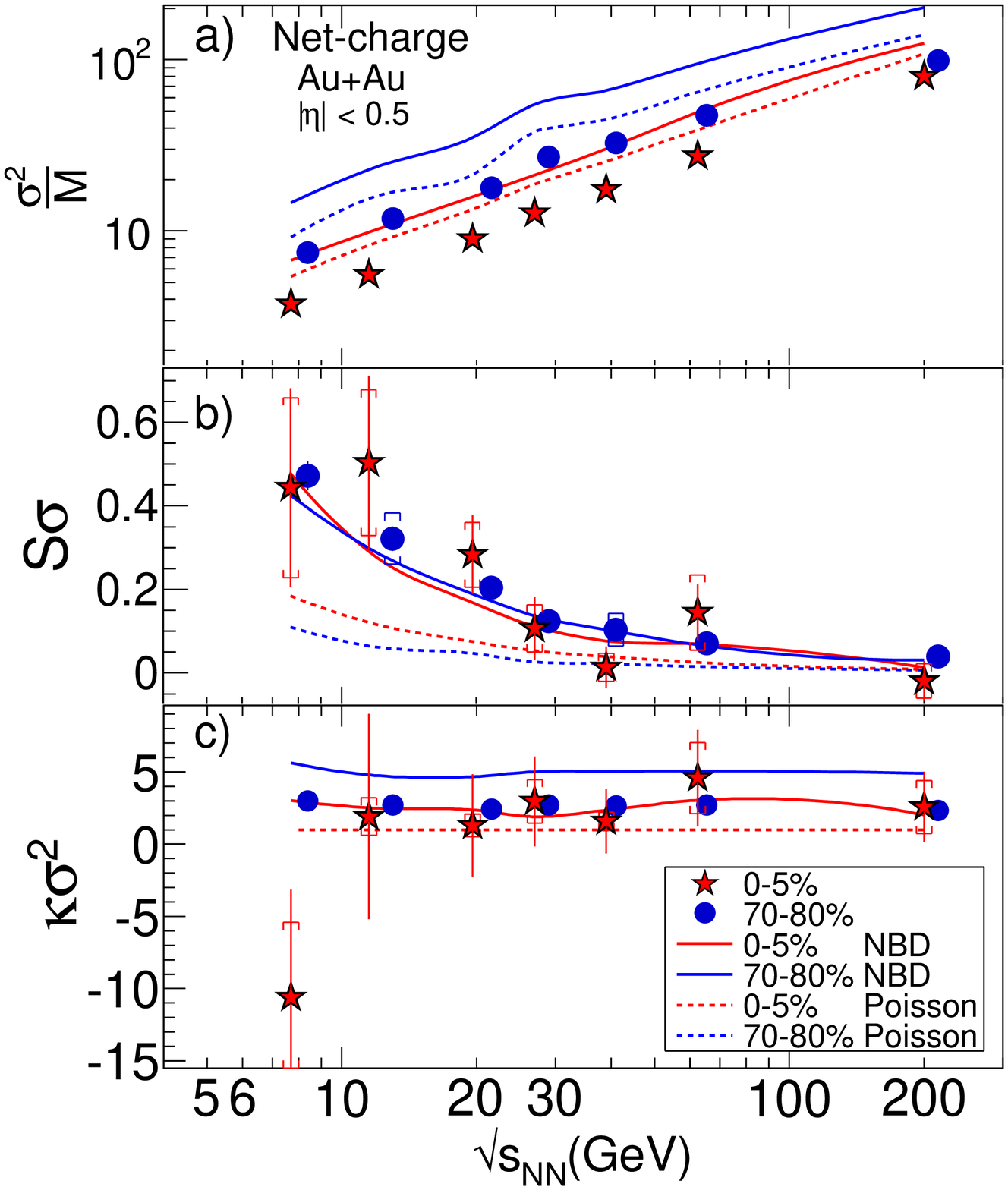}
\caption{\label{star_netcharge}Collision energy dependence of the net-charge 
fluctuations in central collision~\cite{STAR_netcharge}.}
\end{minipage}\hspace{2pc}%
\begin{minipage}{18pc}
\vspace{-2pc}%
\includegraphics[width=18pc,height=18pc]{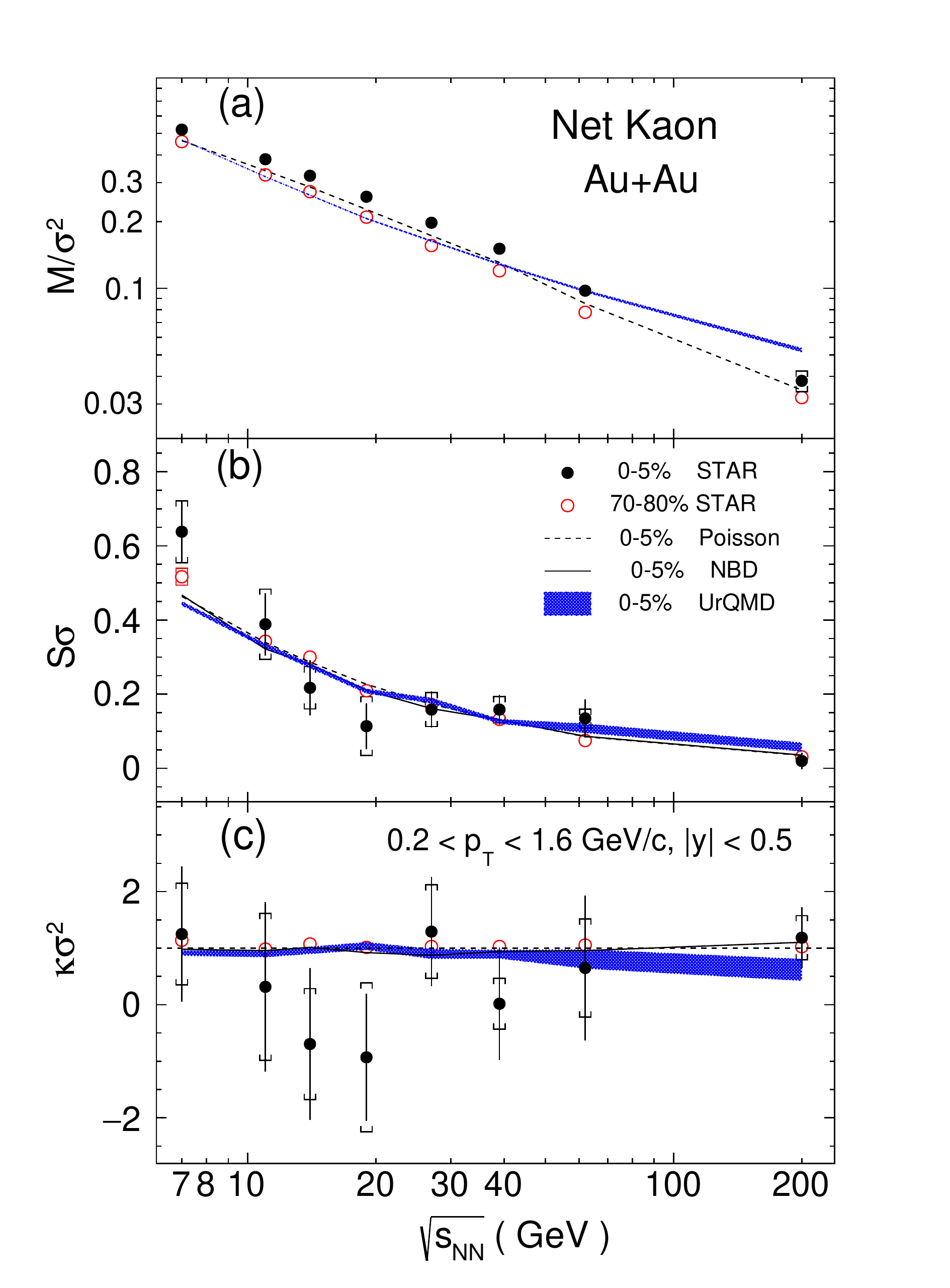}
\caption{\label{star_netkaon}Collision energy dependence of the net-kaon 
fluctuations in central collisions~\cite{STAR_netkaon}.}
\end{minipage} 
\end{center}
\end{figure}
The results of the cumulants of net-charge multiplicity distributions~\cite{STAR_netcharge}
for $|\eta|<0.5$ are shown in Fig.~\ref{star_netcharge}. 
Weak centrality dependence is observed for both \Ss~and \KV~at all 
energies. However, within the present uncertainties, no non-monotonic 
behavior as a function of collision energy. 
The net-kaon results~\cite{STAR_netkaon} are shown in Fig.~\ref{star_netkaon}. The 
collision energy dependence of products of cumulants is seen to be 
smoothly varying as a function of collision energy. No significant 
collision centrality dependence is observed for both  $M/\sigma^2$  and \Ss~at all 
energies. Although the collision centrality and energy dependence of 
\KV~look very intriguing, no definitive statement can be made within 
the current experimental uncertainties. 

\begin{figure}[tbp]
\begin{center}
\includegraphics[width=31pc,height=15pc]{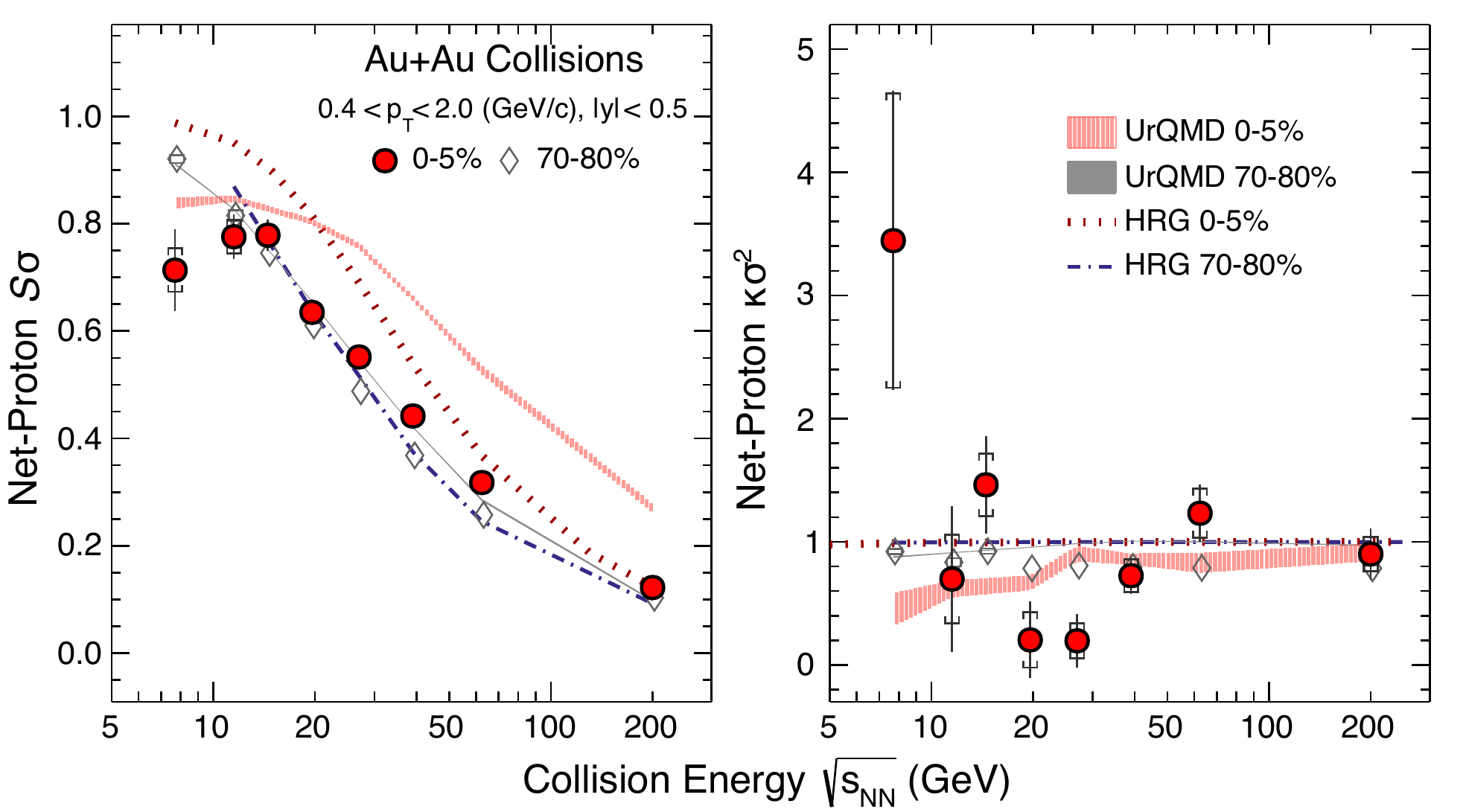}
\caption{\label{netproton}Collision energy dependence of the net-proton 
fluctuations in central and peripheral collisions~\cite{STAR_netproton}.}
\end{center}
\end{figure}
The experimental results of the net-proton multiplicity distributions 
have been reported in 
Ref.~\cite{STAR_netproton,STAR_netproton1,STAR_netproton2}, 
and summarized in Fig.~\ref{netproton}. The collision energy 
dependence of \Ss~shows a smooth variation as a function of energy 
except at very low energies. On the other hand, the nature of 
\KV~variation is very different. For peripheral collisions, there is no 
variation with energy, whereas a non-monotonic variation (with 
3.0$\sigma$ significance) 
with beam energy is observed for 
\KV. This variation could not be explained by hadron resonance gas 
(HRG) calculation and the UrQMD transport model simulation. This 
observation is most likely compatible with the theoretical predictions of the 
critical point. This signature has not been 
observed in other observables so far. 

\begin{figure}[tbp]
\begin{center}
\begin{minipage}{17pc}
\includegraphics[width=15pc,height=17pc]{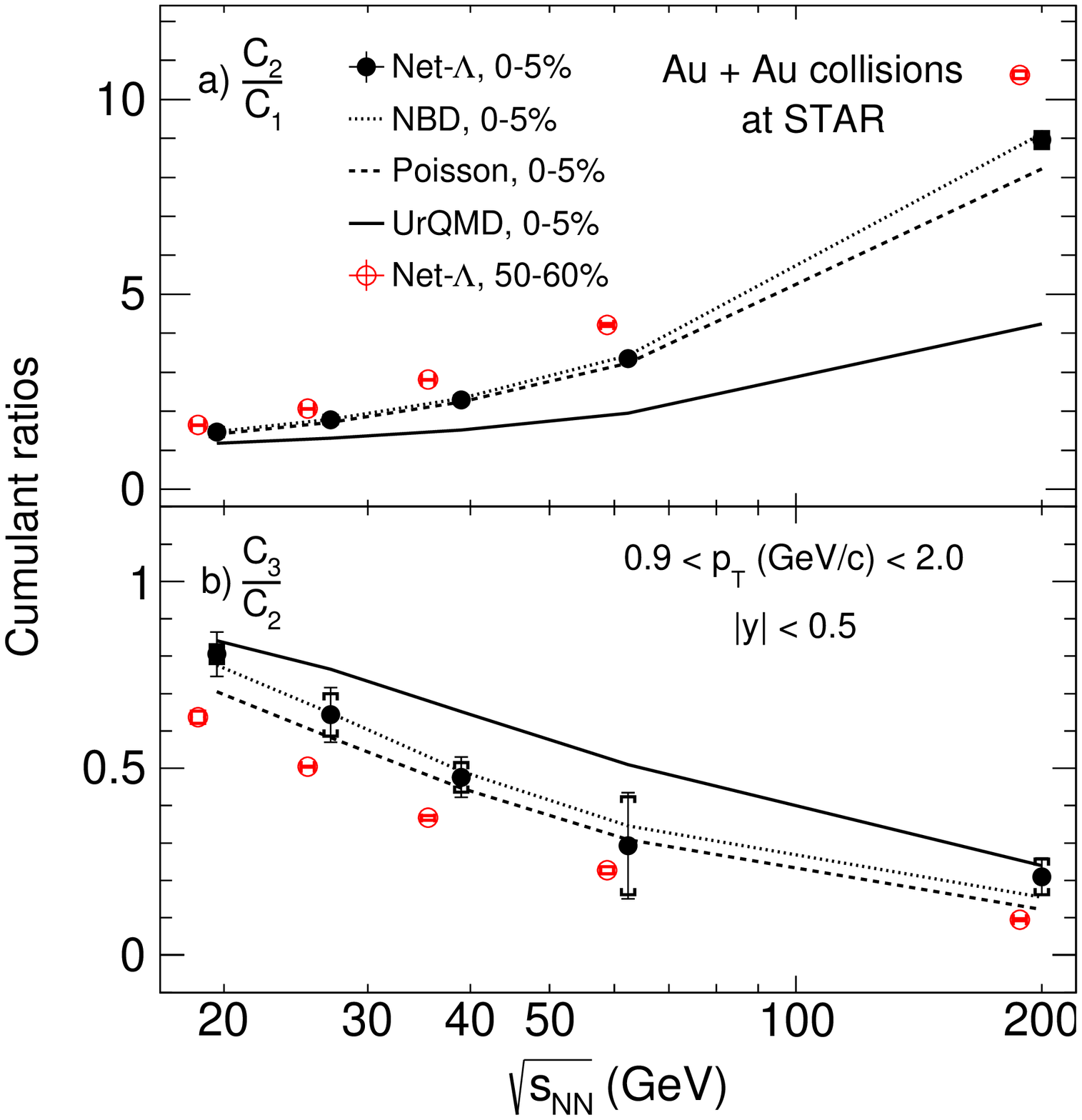}
\caption{\label{netlambda}
Beam energy dependence of  net-$\Lambda$ cumulant ratios,
$C_{2}/C_{1}$, and $C_{3}/C_{2}$ in most central and 
peripheral Au--Au collisions~\cite{STAR_netlambda}.}
\end{minipage}\hspace{2pc}%
\begin{minipage}{17pc}
\includegraphics[width=16.5pc,height=17.5pc]{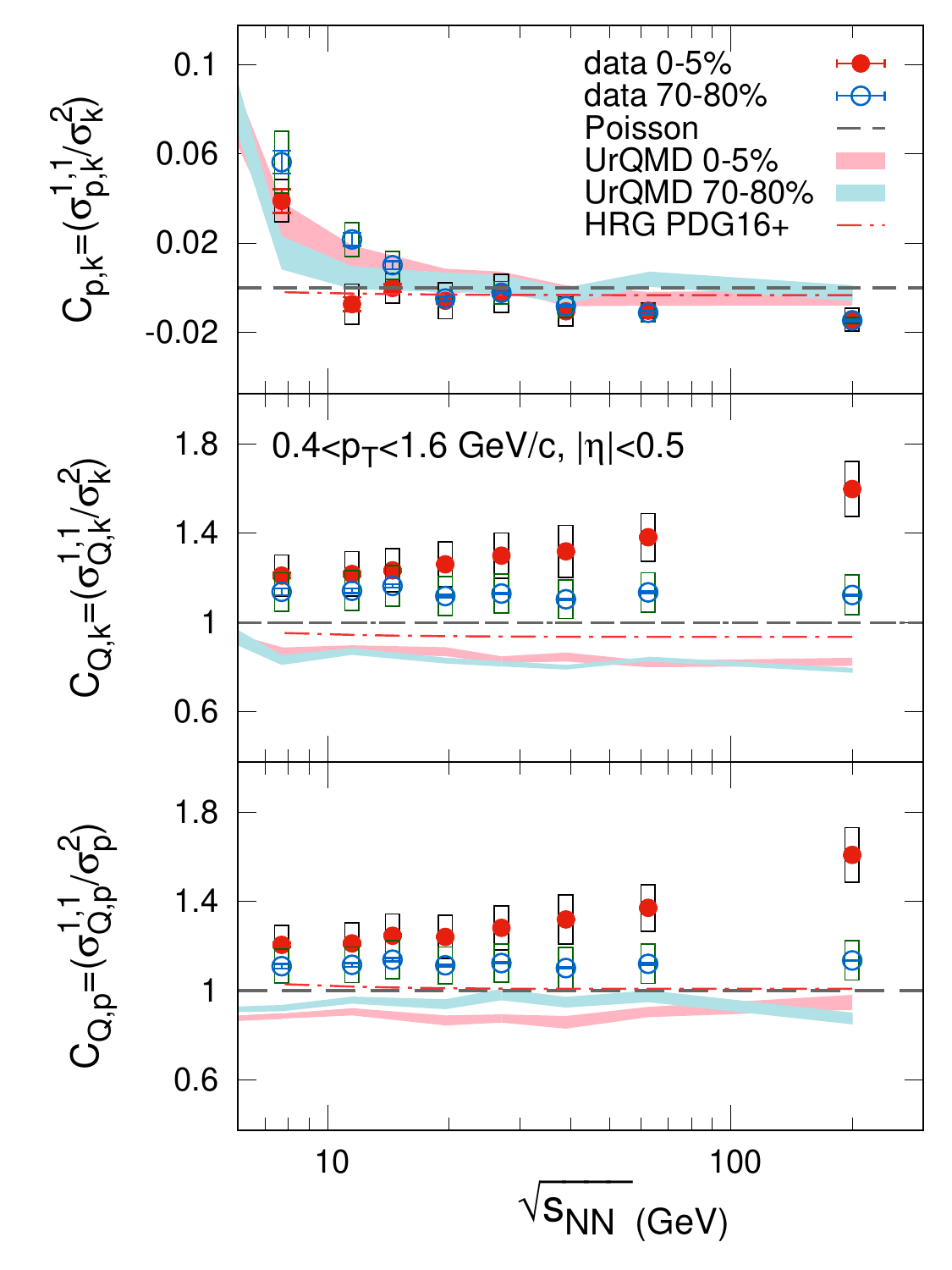}
\caption{\label{offdiagonal}
Beam energy dependence of cumulant ratios 
$C_{\rm p,k}, C_{\rm Q,k}$, and $C_{\rm Q,p}$, of net-proton, net-kaon 
and net-charge for Au--Au collisions~\cite{STAR_offdiagonal}.}
\end{minipage} 
\end{center}
\end{figure}

STAR collaboration has recently reported the beam energy dependence of 
net-$\Lambda$ cumulants~\cite{STAR_netlambda}, which is potentially 
interesting towards our comprehensive understanding of particle 
production mechanisms and their 
correlations as $\Lambda$ carry both baryon and strangeness quantum 
number. Results of this challenging measurement in terms of the beam 
energy dependence of the ratios of first, second and third order 
cumulants are shown in Fig.~\ref{netlambda}, which show no 
non-monotonic behavior for the energies and cumulants studied. These 
results are important for the 
understanding of the freeze-out temperature in the context of both 
baryon number and strangeness conservation. 

\section{Off-diagonal cumulants of conserved quantities}

The diagonal and off-diagonal 
susceptibilities of second order can be expressed in terms of 
second order central moments ($\sigma$):
\begin{center}
$\begin{pmatrix}
  \sigma_{Q}^{2}           & \sigma_{QB}^{1,1} & \sigma_{QS}^{1,1} \\ \\
  \sigma_{BQ}^{1,1}       & \sigma_{B}^{2}     & \sigma_{BS}^{1,1} \\ \\
 \sigma_{SQ}^{1,1}       & \sigma_{SB}^{1,1}  & \sigma_{S}^{2}. 
\end{pmatrix}$
\end{center}
The measurement of these observables give information on the flavor 
carrying susceptibilities. 

The ratios of diagonal and off-diagonal elements are  constructed to cancel the volume effect. 
In the quasiparticle picture of quarks and 
gluons, the ratios $\chi^{11}_{BS}/\chi^{2}_{S}$ and $\chi^{11}_{QS}/\chi^{2}_{S}$ 
are -1/3 and 1/3, respectively. Other ratios, like 
$\chi^{11}_{QB}/\chi^2_B$ has no contribution 
from both light and strange quarks. Thus the following ratios are constructed:
\bea 
 C_{BS} = -3\frac{\chi^{11}_{BS}}{\chi^2_S},  &~~~~&  C_{SB} = -\frac{1}{3}\frac{\chi^{11}_{BS}}{\chi^2_B}, \\
 C_{QS} = 3\frac{\chi^{11}_{QS}}{\chi^2_S},   &~~~~&  C_{SQ} = \frac{\chi^{11}_{QS}}{\chi^2_Q}, \\
 C_{QB} = \frac{\chi^{11}_{QB}}{\chi^2_B},     &~~~~&  C_{BQ} = \frac{\chi^{11}_{QB}}{\chi^2_Q}. 
\eea 
Beam energy dependence of the ratios ($C_{\rm p,k}, C_{\rm Q,k}$, and 
$C_{\rm Q,p}$)~\cite{STAR_offdiagonal} 
are shown in Fig.~\ref{offdiagonal} for central and peripheral 
collisions. The values of $C_{\rm p,k}$ are negative at 200~GeV, and 
change sign around 19.6~GeV for most central collisions. Both 
$C_{\rm Q,p}$ and $C_{\rm Q,k}$ show strong centrality dependence 
indicating the presence of a large excess correlation in central 
events in comparison with peripheral events. 
In addition, the strong dependence of the cumulants is observed 
with the phase space window of measurements. With higher statistics 
datasets and improved acceptance of the STAR detector during the 
second phase of the BES program (BES-II) it will be possible to 
measure higher-order off-diagonal cumulants.

\section{Exploring the formation of DCC domains}

\begin{figure}[h]
\begin{center}
\begin{minipage}{16pc}
\includegraphics[width=16pc,height=14.2pc]{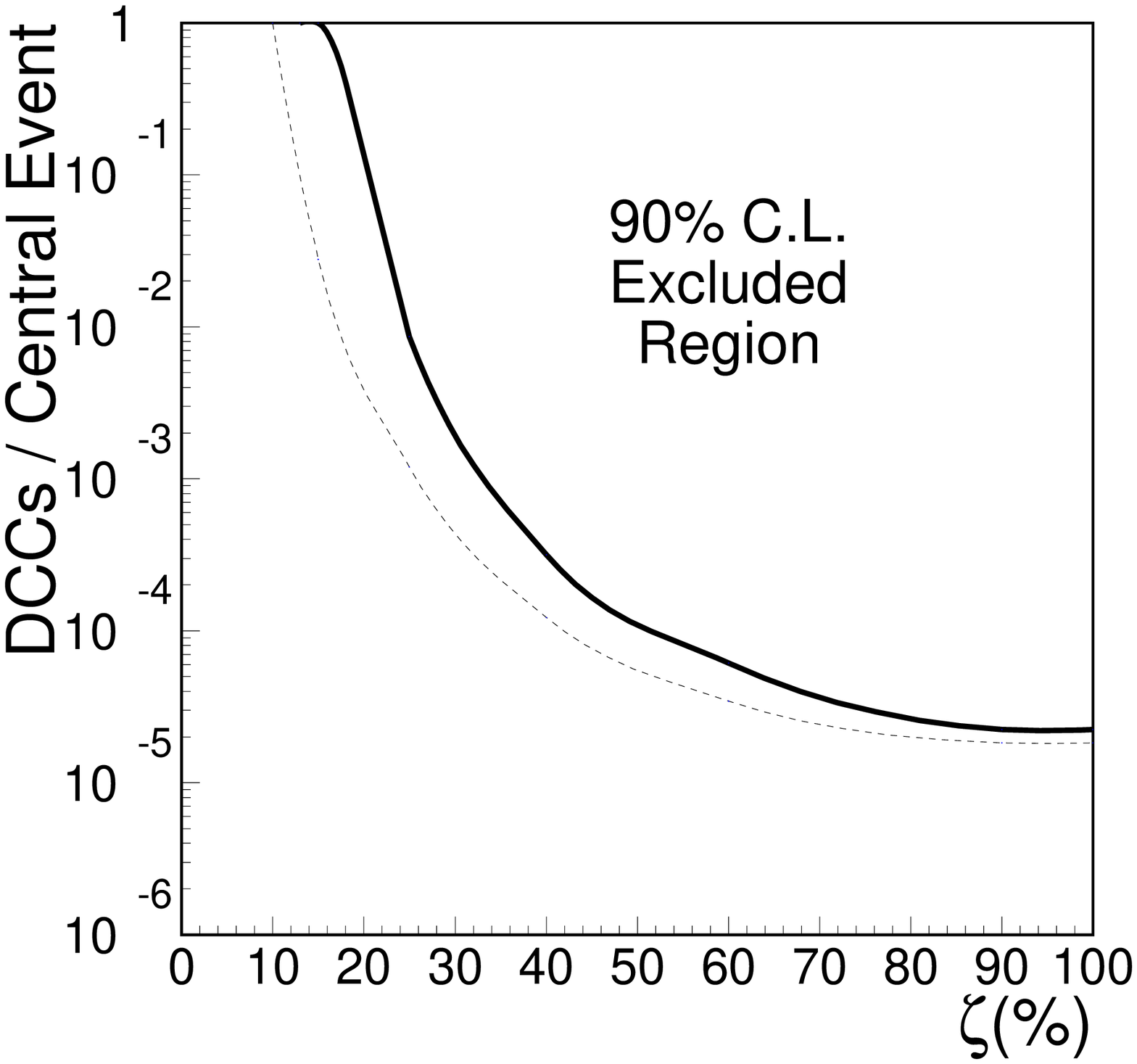}
\caption{\label{wa98_dcc}
90\% C.L 
upper limit on DCC production per central event in 158 AGeV Pb--Pb 
Collisions as a function of 
the fraction of DCC pions\cite{wa98dcc-1}.}
\end{minipage}\hspace{2pc}%
\begin{minipage}{18pc}
\vspace{1.pc}%
\includegraphics[width=18pc,height=14pc]{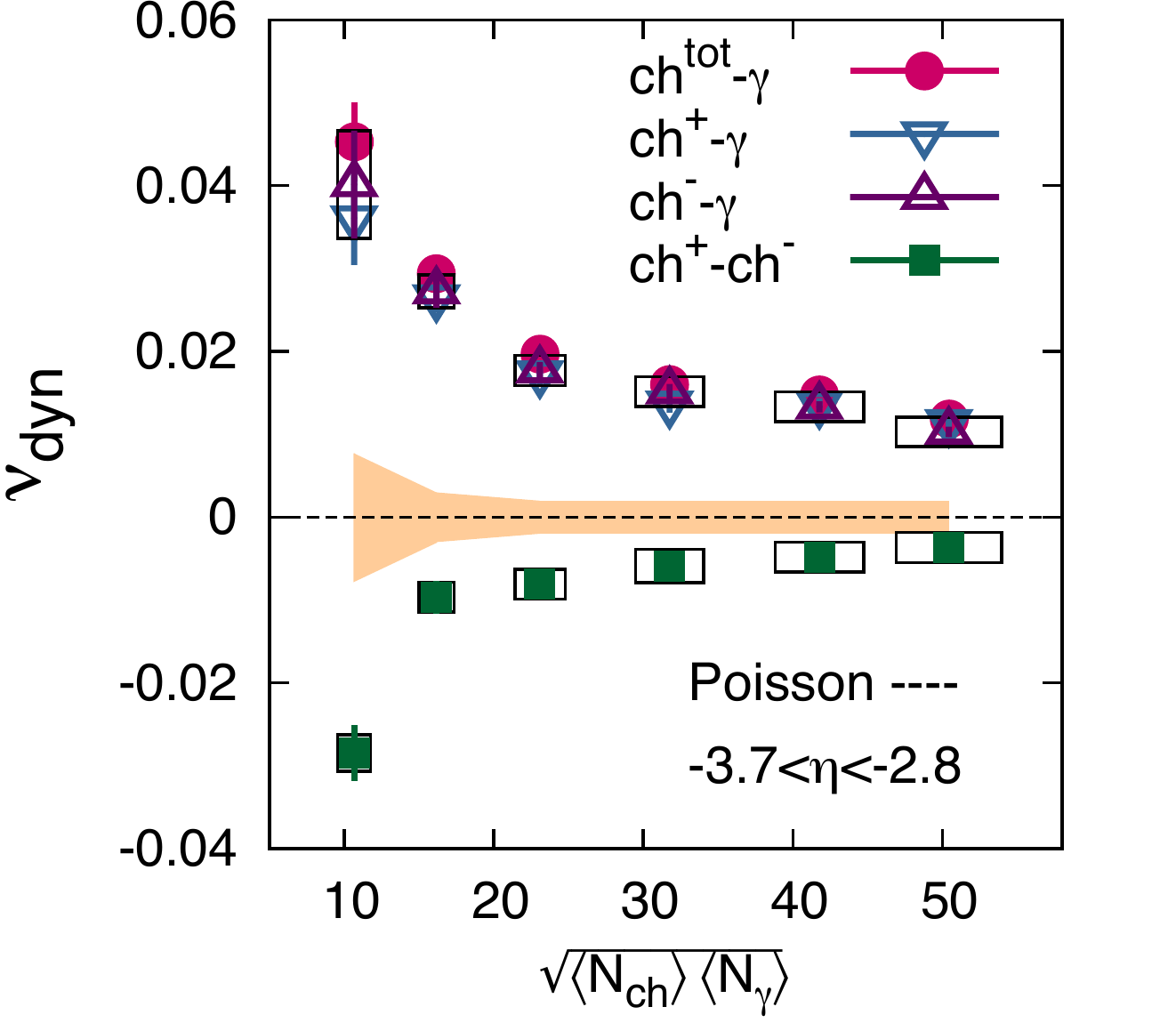}
\caption{\label{star_dcc} 
Dynamical correlation between positive and 
negative charged particles in Au--Au collisions at \sNN=200GeV~\cite{stardcc}. 
}
\end{minipage} 
\end{center}
\end{figure}

The DCC domains are expected to emit pions 
    coherently from the collision volume, resulting in large 
    fluctuations in the fraction of charged to neutral pions. The 
    neutral pion fraction for DCC domains is predicted to follow a 
    probability distribution of the form $P(f) = 1/2\sqrt{f}$, which is 
    different from normal events. Heavy-ion experiments at the 
    CERN SPS~\cite{nayakdcc,wa98dcc-1,wa98dcc-2,wa98dcc-3} have put
    upper limits on the DCC formation as shown Fig~\ref{wa98_dcc}, whereas 
    anomalous fluctuations have been reported at 
    RHIC~\cite{stardcc}. A 
    fresh look at RHIC and LHC energies using both pion and kaon 
    sectors~\cite{ranjitdcc}  is needed to infer about the formation of the DCC domains. 

\section{Mapping the little bang: local fluctuations}

In little bangs, the produced fireball goes through a rapid evolution from an early state of
partonic quark-gluon plasma (QGP) to a hadronic phase
and finally freezes out within a few tens of fm. 
Heavy-ion experiments are predominantly sensitive to the conditions that prevail at the later stage of the collision as
majority of the particles are emitted near the freeze-out.
As a result, a direct and quantitative estimation of the
properties of hot and dense matter in the early stages
and during each stage of the evolution has not yet been possible.   

\begin{figure}[h]
\begin{center}
\includegraphics[width=31pc,height=16pc]{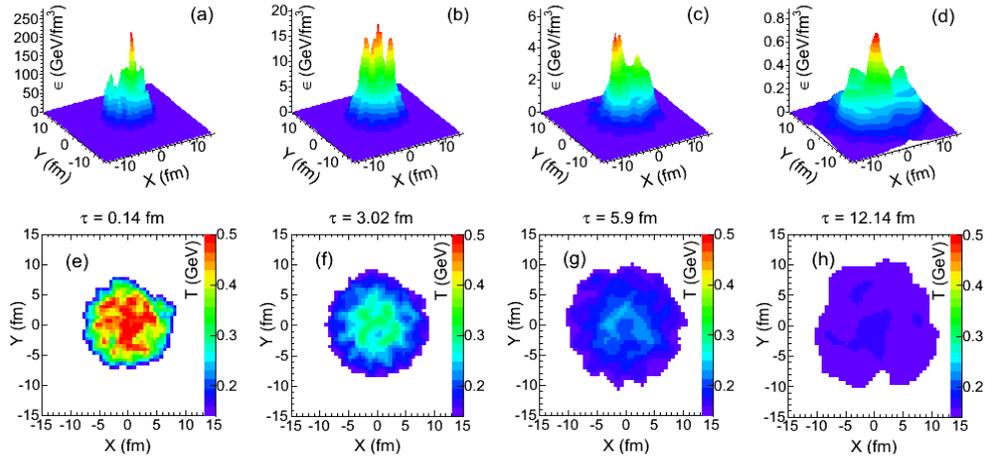}
\caption{\label{hydro_evolution}Hydrodynamic calculations of energy density (upper panels) and temperature 
(lower panels) in the transverse plane for a single 
Pb--Pb event at LHC energy~\cite{evolution_fluctuation}.}
\end{center}
\end{figure}
\begin{figure}[htbp]
\begin{center}
\includegraphics[width=20pc, height=14pc]{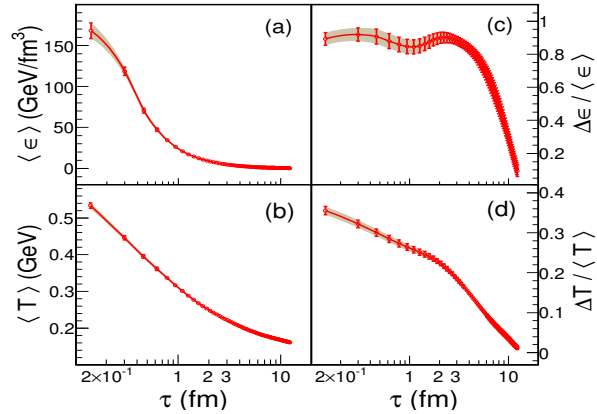}
\caption{\label{fluctuation_evolution}Temporal evolution of (a) average energy density, (b) average temperature, (c) fluctuations in energy density, and (d) fluctuations in temperature for central 
Pb--Pb collisions at LHC energies obtained from hydrodynamic calculations~\cite{evolution_fluctuation}.}
\end{center}
\end{figure}
Relativistic hydrodynamic calculations reveal
fluctuations of initial energy density and temperature, which may
survive till the freeze-out. 
Initial fluctuating conditions have been found to be necessary for
explaining observed elliptic flow in central collisions and
substantial triangular flow of charged
particles~\cite{hannu}. The initial state fluctuations may have
their imprint on the bin to bin local fluctuations within an event.
Event-by-event hydrodynamic calculations provide
a strong theoretical basis for studying the global and local
fluctuations in $\epsilon$ and $T$~\cite{evolution_fluctuation}.
The local fluctuations have been
quantified throughout the evolution by simulating central
Pb--Pb events at LHC energy by the 
use of a (2+1)-dimensional event-by-event ideal hydrodynamical
framework with lattice-based EOS~\cite{hannu}. 

In Fig.~\ref{hydro_evolution}, we present distributions of $\epsilon$ and
$T$ in the transverse plane at four proper times ($\tau$). 
At early times, sharp and pronounced peaks in
$\epsilon$ and hotspots in $T$ are observed.
Large bin-to-bin fluctuations observed in $\epsilon$ and
$T$ indicate that the system formed immediately after collision is quite
inhomogeneous in phase space. As time elapses, the system
cools, expands, and the bin-to-bin variations in $\epsilon$ and $T$
become smooth. These observations are quantified in terms of the
average over all the bins and bin to bin fluctuations in $\epsilon$
and $T$, plotted as a function of $\tau$, shown in
Fig.~\ref{fluctuation_evolution}.
We observe that $\epsilon$~decreases sharply up to $\tau=1$~fm, and
then the decrease is slow till freeze-out.
The fall of $T$~with $\tau$ is rather smooth. 
At early times, the fluctuations 
are observed to be very large, and then decrease rapidly. It is thus clear that 
a detailed insight into the evolution of fluctuations is possible by
studying local fluctuations in $\epsilon$ and $T$. 

\section{Summary and outlook} 

In this article, we have discussed a number of fluctuation techniques
for understanding the nature of the QCD phase transition over a wide range of baryon
chemical potential as well as to locate the critical
point. One of the first fluctuation studies is in terms of event-by-event fluctuations
of higher-order cumulants of conserved quantities, both diagonal and
off-diagonal elements. A comparison with lattice and HRG model
calculations provide measures of the freeze-out conditions.
Assuming that the signal at freeze-out survives dissipation
during the evolution of the fireball from the hadronization stage, the
higher cumulants can be used as one of the preferred tools for locating
the critical point. The products of the cumulants are observed to have smooth variation as a function of
collision energy for net-charge, net-kaon, and net-$\Lambda$ distributions.
But the experimental results of \KV~at RHIC energies show a sign of non-monotonic
behavior in the net-proton multiplicity distributions. 
We look forward to getting the confirmation of the critical point with future 
higher statistics data for higher order cumulants of conserved 
quantities as well as more differential measurements 
in rapidity and \pT.

We have discussed fluctuations in particle multiplicity,
mean transverse momentum ($\langle p_{\rm T} \rangle$) and temperature
to extract isothermal compressibility, specific heat, and the speed of
sound. These observables, being sensitive to the phase transition,
provide important measures for the nature of the transition and to
locate the critical point. Charge-neutral fluctuations in the pion and
kaon sectors are discussed in terms of providing signatures of the formation
of disoriented chiral condensates. Another topic in terms
of fluctuation is to construct local fluctuation maps in rapidity and
azimuthal bins. By making a
correspondence of measured fluctuations with the time evolution
of the fluctuations from theoretical calculations, it is possible to
infer the thermodynamic conditions at different stages
of the QGP evolution. 
Most of the fluctuation studies require large coverage of the
detectors and large statistics measurements. The STAR experiment with
RHIC BES-II is ideal for the search of the critical point. 
The next-generation multipurpose detector at the LHC as a follow-up to
the present ALICE experiment will have a large coverage, which will be
suitable for event-by-event fluctuations~\cite{ALICE_new} to probe
critical fluctuations at \muB=0.

\smallskip

\end{document}